\documentclass[10pt, superscriptaddress, reprint, amsmath, amssymb, aps, pra, showpacs]{revtex4-2}

\usepackage{amsmath,amssymb,mathrsfs}
\usepackage{cases}
\usepackage{graphicx}
\usepackage{dcolumn}
\usepackage{bm}
\usepackage{color}
\usepackage{soul}
\usepackage{xcolor}

\usepackage{amsfonts,latexsym}
\usepackage{enumerate}
\usepackage{rotating}
\usepackage{epstopdf}
\usepackage{braket}
\usepackage{MnSymbol,bbding,pifont}
\usepackage{makecell} 

\usepackage{hyperref}
\hypersetup{
	colorlinks=true,
	linkcolor=blue,
	filecolor=blue,
	urlcolor=blue,
	citecolor=magenta
}

\newcommand{\beq}{\begin{equation}}
	\newcommand{\eeq}{\end{equation}}
\newcommand{\beqa}{\begin{eqnarray}}
	\newcommand{\eeqa}{\end{eqnarray}}
\newcommand{\Beqa}{\begin{eqnarray*}}
	\newcommand{\Eeqa}{\end{eqnarray*}}

\def\bal#1\eal{\begin{align}#1\end{align}}
\def\Bal#1\Eal{\begin{align*}#1\end{align*}}

\newcommand{\mi}{\mathrm{i}}

\begin{document}

\title{Tunnel-coupled optical microtraps for ultracold atoms}

\author{Shangguo Zhu}	
\author{Yun Long}
\affiliation{State Key Laboratory of Optical Technologies on Nano-Fabrication and Micro-Engineering, Institute of Optics and Electronics, Chinese Academy of Sciences, Chengdu 610209, China} 
\affiliation{National Key Laboratory of Optical Field Manipulation Science and Technology, Chinese Academy of Sciences, Chengdu 610209, China} 
\affiliation{Research Center on Vector Optical Fields, Institute of Optics and Electronics, Chinese Academy of Sciences, Chengdu 610209, China} 
\affiliation{School of Optoelectronics, University of Chinese Academy of Sciences, Beijing 100049, China} 

\author{Wei Gou}
\affiliation{State Key Laboratory of Optical Technologies on Nano-Fabrication and Micro-Engineering, Institute of Optics and Electronics, Chinese Academy of Sciences, Chengdu 610209, China} 
\affiliation{National Key Laboratory of Optical Field Manipulation Science and Technology, Chinese Academy of Sciences, Chengdu 610209, China} 
\affiliation{Research Center on Vector Optical Fields, Institute of Optics and Electronics, Chinese Academy of Sciences, Chengdu 610209, China} 

\author{Mingbo Pu}	
\affiliation{State Key Laboratory of Optical Technologies on Nano-Fabrication and Micro-Engineering, Institute of Optics and Electronics, Chinese Academy of Sciences, Chengdu 610209, China} 
\affiliation{National Key Laboratory of Optical Field Manipulation Science and Technology, Chinese Academy of Sciences, Chengdu 610209, China} 
\affiliation{Research Center on Vector Optical Fields, Institute of Optics and Electronics, Chinese Academy of Sciences, Chengdu 610209, China} 
\affiliation{School of Optoelectronics, University of Chinese Academy of Sciences, Beijing 100049, China}

\author{Xiangang Luo}
\email[]{lxg@ioe.ac.cn}
\affiliation{State Key Laboratory of Optical Technologies on Nano-Fabrication and Micro-Engineering, Institute of Optics and Electronics, Chinese Academy of Sciences, Chengdu 610209, China} 
\affiliation{National Key Laboratory of Optical Field Manipulation Science and Technology, Chinese Academy of Sciences, Chengdu 610209, China} 
\affiliation{School of Optoelectronics, University of Chinese Academy of Sciences, Beijing 100049, China}

\begin{abstract}	
Arrays of individual atoms trapped in optical microtraps with micrometer-scale sizes have emerged as a fundamental, versatile, and powerful platform for quantum sciences and technologies.
This platform enables the bottom-up engineering of quantum systems, offering the capability of low-entropy preparation of quantum states with flexible geometry, as well as manipulation and detection at the single-site level. 
The utilization of ultracold itinerant atoms with tunnel coupling in optical microtraps provides new opportunities for quantum simulation, enabling the exploration of exotic quantum states, phases, and dynamics, which would otherwise be challenging to achieve in conventional optical lattices due to high entropy and limited geometric flexibility.  
Here, the development of tunnel-coupled optical microtraps for the manipulation of ultracold atomic quantum systems and its recent advances are briefly reviewed.
\end{abstract}

\maketitle

\section{Introduction}
\label{sect:intro}

Optical trapping has been an essential part of the toolbox in quantum science and technologies enabling the confinement, cooling, and detection of ultracold atomic quantum gases. 
Since the first realization of Bose-Einstein condensates (BECs) of alkali atoms\cite{Anderson1995, Davis1995} in 1995, multiple optical microtraps of micrometer-scale size have provided a rich playground for studying quantum interference~\cite{Andrews1997}, quantum dynamics and transport of superfluids~\cite{Albiez2005, Valtolina2015}. 
Later in 2001, it was demonstrated~\cite{Schlosser2001} that single atoms could be loaded to optical microtraps and read out with high-fidelity fluorescence imaging.  
These optical microtraps, often referred to as ``optical tweezers", are created using tightly focused beams of light generated with a high-numerical-aperture (NA, typically ${\rm NA} \ge 0.5$) objective lens.

In the past two decades, the advent of the second quantum revolution, characterized by the precise control of individual quantum systems, has sparked significant interest in the optical tweezers for the manipulation of individual ultracold atoms.  
Unlike the well-established top-down approach of loading many atoms into optical lattices~\cite{Bloch2008, Eckardt2017, Gross2017, Gross2021, Schafer2020}, optical tweezers provide a bottom-up method for assembling quantum systems, offering significant advantages in terms of flexibility in geometry and spatial arrangement~\cite{Zimmermann2011, Endres2016, Barredo2016, Barredo2018, Wang2020}, and precise control at the single-site level.
In addition, the optical tweezers allow for low-entropy preparation of quantum states. 
The motional entropy in tweezers can be eliminated by deterministic atom number preparation~\cite{Serwane2011} and cooling the individual atoms to the ground state of each tweezer through sideband cooling techniques~\cite{Kaufman2012, Thompson2013}. 
Another major source of entropy associated with probabilistic trap occupation could be effectively eliminated by employing gray-molasses loading techniques with enhanced loading efficiency~\cite{Grunzweig2010, Lester2015, Brown2019a, Aliyu2021, Jenkins2022, Angonga2022}, atom rearrangement to form defect-free arrays via extra mobile tweezers~\cite{Endres2016, Barredo2016}, and postselection of the images~\cite{Spar2022, Yan2022}. 
With these unique advantages, over the last decade, individual atom arrays trapped in optical tweezers have emerged as a prominent and versatile platform for quantum sciences and technologies, in fields including quantum computation~\cite{Levine2019, Madjarov2020, Jenkins2022, Ma2022, Graham2022, Bluvstein2022}, quantum simulation~\cite{Bernien2017, Lienhard2018, Omran2019, De_Leseleuc2019, Keesling2019, Browaeys2020, Kaufman2021, Ebadi2021, Scholl2021, Semeghini2021}, quantum metrology~\cite{Madjarov2019, Norcia2019, Young2020}, and quantum information~\cite{Darquie2005, Beugnon2006, Saffman2010}.

The phenomenon of tunneling through a potential barrier has been a longstanding and fundamental topic in physics. 
It finds diverse applications, including the tunneling of electrons in scanning tunneling microscopy~\cite{Binnig1987} and the coupling of evanescent light fields in catenary optics~\cite{Pu2015, Luo2019, Pu2018, Luo2004}. 
The macroscopic quantum tunneling dynamics of BECs has been experimentally observed in double-well optical traps~\cite{Albiez2005}. 
Additionally, the tunneling of ultracold atoms between neighboring trapping sites in optical lattices leads to the tunnel coupling, a natural realization of the tunneling matrix element in the Hubbard model~\cite{Jaksch1998, Greiner2002}. 
The combination of the unique advantages offered by the tweezer platform and the presence of light-mass itinerant atoms, such as lithium atoms, in tunnel-coupled optical tweezer arrays has garnered significant interest~\cite{Kaufman2014, Murmann2015, Spar2022, Yan2022, Kaufman2021, Gross2021}.  
In contrast to Rydberg atoms with strong van der Waals interactions or polar molecules with dipolar interactions, itinerant atoms in tunnel-coupled optical tweezer arrays present new opportunities for directly simulating itinerant lattice models and exploring quantum states, phases, and dynamics involving tunnel coupling, including the study of Hubbard models and quantum walks. 
This allows for the investigation of fundamental quantum phenomena with enhanced control and flexibility, potentially unlocking novel applications in quantum sciences and technologies.

In this review article, we focus on the development and recent advances in tunnel-coupled optical microtraps as a fundamental tool for the manipulation of ultracold atomic quantum gases. 
In Sect.~\ref{sect:dw}, we begin by briefly introducing the early applications of tunnel-coupled double-well optical microtraps, which were used in studying the tunneling dynamics of cold atomic superfluids.
In Sect.~\ref{sect:tweezer}, We then focus on the utilization of tunnel-coupled optical tweezers for the precise manipulation of ultracold individual atoms. 
In Sect.~\ref{sect:summary}, we provide a summary and future prospects in these areas of research. By examining these topics, we aim to provide a comprehensive overview of the progress and potential directions in the field of tunnel-coupled optical microtraps for ultracold atomic quantum gases.

\section{Superfluids in tunnel-coupled double wells} 
\label{sect:dw}

The use of tunnel-coupled double wells (DWs) formed by micrometer-scale optical microtraps to trap BECs has served as an early paradigm for quantum simulation~\cite{Georgescu2014, Gross2017}, allowing for the emulation and investigation of the fascinating properties exhibited by supercurrents in solid-state superconducting systems, such as the atomic Josephson effect~\cite{Josephson1962}.

\textbf{Tunneling dynamics and atomic Josephson effect.}
The quantum tunneling dynamics of BECs in tunnel-coupled DWs has been extensively studied in early theoretical investigations~\cite{Javanainen1986, Jack1996, Zapata1998, Milburn1997, Smerzi1997, Raghavan1999, Giovanazzi2000, Meier2001, Sakellari2002, Shchesnovich2004, Wang2008, Mayteevarunyoo2008}. 
The DWs serve as a beam splitter for the matter wave of BECs, enabling the experimental measurement of the macroscopic quantum phases between the BECs in the two wells via matter-wave interferometry~\cite{Andrews1997, Saba2005}. 
Also, with the tunneling barrier, BECs in tunnel-coupled DWs naturally emulate the behavior of Josephson junctions (JJs) found in solid-state superconducting systems, forming bosonic JJs.

\begin{figure*}[ht!]
	\centering
	\includegraphics[width=0.98\linewidth]{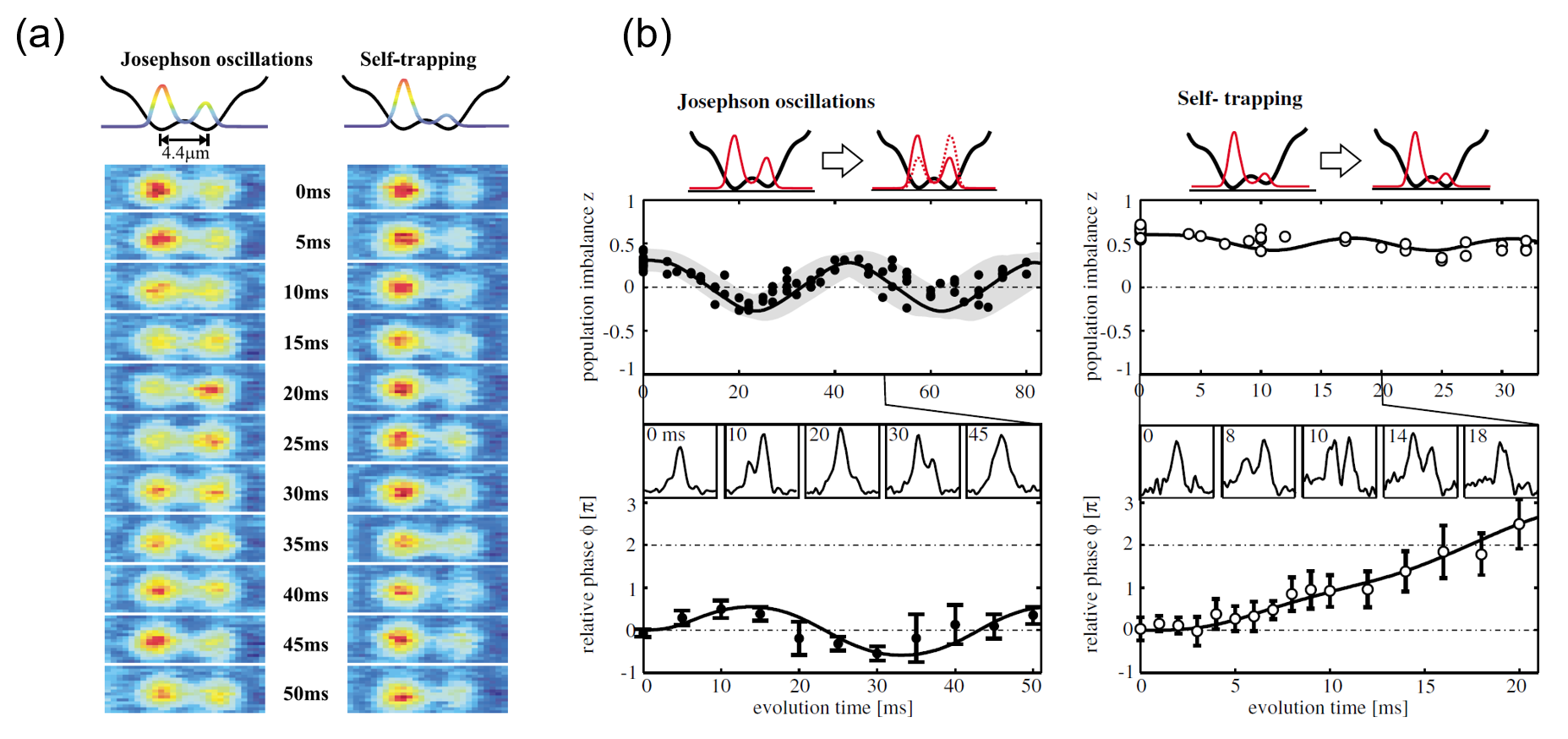}
	\caption{
		Experimental observation of Josephson oscillations and macroscopic quantum self-trapping effect of a BEC in a symmetric double-well potential. 
		\textbf{(a)} The relative population in the double wells over time. In the left column (Josephson oscillations), the oscillation of population is observed when the initial population difference is set to be below a critical value. In the right panel (Self-trapping), when the initial population difference exceeds the critical value, the relative population does not change over time, exhibiting a macroscopic quantum self-trapping effect. 
		\textbf{(b)} The detailed time dependence of the relative population difference $z = (N_L - N_R) / (N_L + N_R)$ and the quantum phase difference $\phi = \phi_R - \phi_L$, where $N_{L, R}$ and $\phi_{L, R}$ are the atom number and the quantum phase of the left ($L$) and right ($R$) components, respectively. 
		Adapted with permission.\textsuperscript{\cite{Albiez2005}} Copyright 2005, American Physical Society. }
	\label{fig:Albiez2005}
\end{figure*}

In experimental studies with a BEC of $^{87}$Rb atoms in DWs, created by the superposition of an optical trap with a strong harmonic confinement, the phenomena of Josephson oscillations and macroscopic quantum self-trapping effect were observed~\cite{Albiez2005} (see Figure~\ref{fig:Albiez2005}). 
These observations provide experimental evidence for the coherent tunneling dynamics of the BEC between the two wells and the interplay between tunneling and interactions in the system. 
The Josephson oscillations manifest as periodic oscillations of the population imbalance between the two wells, while the counterintuitive quantum self-trapping effect corresponds to the stabilization of the population imbalance at certain values due to the nonlinearity of the interatomic interactions.

The dynamics of the BECs can be described by two coupled differential equations derived~\cite{Smerzi1997} from the Gross-Pitaevskii equation
\begin{align}
	\dot{z} &= -\sqrt{1-z^2}\sin \phi, \label{eq:JJeq1}\\
	\dot{\phi} &= \Lambda z + \frac{z}{\sqrt{1-z^2}} \cos \phi, \label{eq:JJeq2}
\end{align}
where $\Lambda$ is a parameter proportional to the ratio of the on-site interaction energy and the coupling matrix element~\cite{Smerzi1997}, the relative population difference $z = (N_L - N_R) / (N_L + N_R)$, the quantum phase difference $\phi = \phi_R - \phi_L$, and $N_{L, R}$ and $\phi_{L, R}$ are the atom number and the quantum phase of the left ($L$) and right ($R$) components, respectively. 
When the initial population difference $z(0)$ is small, the two coupled equations can be approximated as $\ddot{\phi} \propto - \phi$. This resembles the classical equation of motion of a simple pendulum, resulting in an oscillatory solution of the quantum phase $\phi$, known as Josephson oscillation (see the left figures in Figure~\ref{fig:Albiez2005} (a) and (b)).  
When the initial population difference $z(0)$ is large and exceeds a critical value, the relative phase $\phi$ increases rapidly in time, as shown in Equation~(\ref{eq:JJeq2}). 
This leads to a rapidly alternating tunneling current, according to the right-hand side of Equation~(\ref{eq:JJeq1}). 
Then, the population difference $z$ performs small oscillations around its initial value and does not change over time, which is known as the macroscopic quantum self-trapping effect (see the right figures in Figure~\ref{fig:Albiez2005} (a) and (b)).

The study of tunneling dynamics in DWs was extended to ultracold Fermi gases. By using optical traps with a repulsive optical barrier, researchers have investigated the behavior of ultracold Fermi gases of $^6$Li in the context of fermionic JJs across the BEC-BCS crossover regime~\cite{Valtolina2015, Kwon2020}. 
Furthermore, homogeneous ultracold Fermi gases of $^6$Li confined in box traps were employed to realize fermionic JJs in 2D~\cite{Luick2020}.

In addition to the use of tunnel-coupled optical microtraps, DWs can also be implemented in other experimental setups, such as DW arrays in an optical superlattice~\cite{Sebby-Strabley2006}, allowing the observation and control of the second-order superexchange interaction~\cite{Folling2007, Trotzky2008}, the study of quantum phase transitions with parity-symmetry breaking and hysteresis~\cite{Trenkwalder2016}, and the exploration of atomic spin entanglement~\cite{Dai2016}. 
DWs could also be generated in the form of magnetic traps, enabling the observation of alternating-current and direct-current Josephson effects~\cite{Levy2007}, the exploration of the crossover from Josephson dynamics to hydrodynamic behavior~\cite{LeBlanc2011}, and the simulation of sine-Gordon equation~\cite{Coleman1975, Gritsev2007, Langen2015, Schweigler2017, Schweigler2021, Ji2022}.

\section{Individual atoms in tunnel-coupled optical tweezers}
\label{sect:tweezer}

In contrast to the top-down way of loading many cold atoms in optical lattices, optical tweezers provide a bottom-up way to assemble quantum systems from individual atoms trapped one at each tweezer.
This provides the following multiple unique advantages for engineering the desired quantum states.

First, the optical tweezer arrays could in principle form arbitrary patterns, which are usually generated by acousto-optic deflectors and spatial light modulators~\cite{Zimmermann2011, Endres2016, Barredo2016, Barredo2018, Wang2020}. 
The defect-free optical tweezer arrays generated in 1D~\cite{Endres2016}, 2D~\cite{Barredo2016}, and 3D~\cite{Barredo2018} have been demonstrated. 
This provides huge flexibility in the lattice geometry and spatial arrangement for the simulation of quantum many-body states.

Second, optical tweezers allow the preparation of quantum states with very low entropy. 
One major source of entropy, the motional entropy of atoms in the trap could be eliminated by cooling the individual atoms to the ground state of each tweezer~\cite{Serwane2011, Kaufman2012}. 
Another source of entropy associated with probabilistic trap occupation could be effectively eliminated by gray-molasses loading with enhanced loading efficiency~\cite{Grunzweig2010, Lester2015, Brown2019a, Aliyu2021, Jenkins2022, Angonga2022} and atom rearrangement to form defect-free arrays via extra mobile tweezers~\cite{Endres2016, Barredo2016}. 
With the recently developed atom rearrangement algorithms~\cite{Lee2017, Kim2019, Schymik2020, Sheng2021, Ebadi2021, Sheng2022, Singh2022, Cimring2022, Zhu2022, Wang2023, Tian2023}, the rearrangement time cost and success rate could be largely improved, allowing the preparation of large-size defect-free atom arrays. 
The use of cygeonic environment, the vacuum lifetime of a trapped single atom could be further increased to $\approx 6000$ s~\cite{Schymik2021}, theoretically allowing the expansion of array size to $10^4$~\cite{Wang2023}, comparable to the scalability of optical lattices. 
For itinerant atoms in tunnel-coupled tweezer arrays, the entropy could even be reduced to zero via post-selection of the images~\cite{Spar2022, Yan2022}.

Third, the optical tweezers provide the precise control of the atoms at the single-site level. 
Many model parameters like interaction, energy offset, tunneling energy could be precisely tuned for desired sites allowing the study of dynamics in the staggered lattices ~\cite{Spar2022, Yan2022}. 
Combined with the quantum gas microscopes, the images with single-atom resolution provide precise information such as spectral functions and spin correlations~\cite{Gross2021}. 
With these advantages, optical tweezers have emerged as a versatile platform for quantum sciences and technologies.

Here, we list the advantages in the different aspects of optical tweezers and optical lattices in Table~\ref{table:OTOL}. 
Besides the listed aspects, different platforms exhibit different homogeneity features. 
The spacing between neighboring sites in optical lattices is about hundreds of nanometers, usually half the wavelength of the trapping laser; while the spacing in optical tweezer arrays is as large as $1\sim 2$ micrometers. 
As the tunnel coupling is very sensitive to the spacing, the tunnel coupling in optical tweezer arrays is usually much weaker than that in optical lattices. 
Then, to reach the regime where the tunneling is larger than disorder, the intensities of the tunnel-coupled tweezers need to be balanced to be better than $0.5\%$ of the trapping depth~\cite{Spar2022}. 
For Rydberg atoms in tweezer arrays, because the interaction is stronger, this homogeneity requirement is within a few percent~\cite{Endres2016}. 
For optical lattices, there is usually an overall harmonic confinement due to the Gaussian beam intensity profile. 
The lattice sites in the center of the confinement is deeper than those on the edge. 
This hinders the simulation of quantum systems, requiring overall uniformity, in optical lattices.

\begin{table*}[ht!]	
	\renewcommand{\arraystretch}{1.5}
	\centering
	\caption{\label{table:OTOL} Comparison between the advantages of optical tweezers and optical lattices.}
	\begin{tabular}{@{\hskip 0.2cm}l@{\hskip 0.5cm}c@{\hskip 0.5cm}c@{\hskip 0.5cm}c@{\hskip 0.5cm}c@{\hskip 0.2cm}}
		\Xhline{2\arrayrulewidth}
		& Arbitrary Geometry	& Single-site control & Single-site readout & Scalability \\
		\hline
		Optical Lattices  & No					& No  				  & Yes & High       \\
		Optical Tweezers  & Yes					& Yes 				  & Yes & Moderate   \\
		\Xhline{2\arrayrulewidth}
	\end{tabular}
\end{table*}

Usually based on the types of interactions, there could be three major types of atoms trapped in optical tweezers: 1) Rydberg atoms with a strong $1/r^6$ van der Waals interaction enabling the Rydberg blockade, which prevents two nearby atoms from having Rydberg excitations at the same time~\cite{Browaeys2020}; 2) Polar molecules with strong electric dipole-dipole interaction~\cite{Anderegg2019, Cairncross2021, Kaufman2021, Burchesky2021, Christakis2023, Holland2022}; 3) Itinerant atoms with tunnel-coupled interactions~\cite{Spar2022, Yan2022}. 
The itinerant atoms in a large-scale tunnel-coupled optical tweezer array could unlock the direct simulation of Fermi-Hubbard models with low-entropy preparation of quantum states. 
This provides a feasible route to resolve the mechanism of high-temperature superconductors through quantum simulation based on ultracold atoms. 
Recently, the advances in tunnel-coupled optical tweezers have already attracted significant interest. 
Here, we focus on the review of individual ultracold atoms in tunnel-coupled optical tweezers.

\subsection{Development of optical tweezers}

In 2001, Schlosser et al.~\cite{Schlosser2001} demonstrated that single atoms could be loaded to optical tweezers (micrometer-scale foci of light) and detected with high-fidelity fluorescence imaging. 
Laser-cooled atoms in optical tweezers experience collisional blockade~\cite{Schlosser2002}, where atoms are ejected in pairs out of the tweezer through optically enhanced dipole–dipole interactions, resulting in the stochastic sub-Poissonian loading of zero atoms or one atom. 
The development of single-atom sources facilitated seminal experiments such as quantum state control of individual neutral atoms in tweezers~\cite{Jones2007} and the transportation of single atoms using mobile tweezers~\cite{Beugnon2007}.

\begin{figure*}[ht!]
	\centering
	\includegraphics[width=0.98\linewidth]{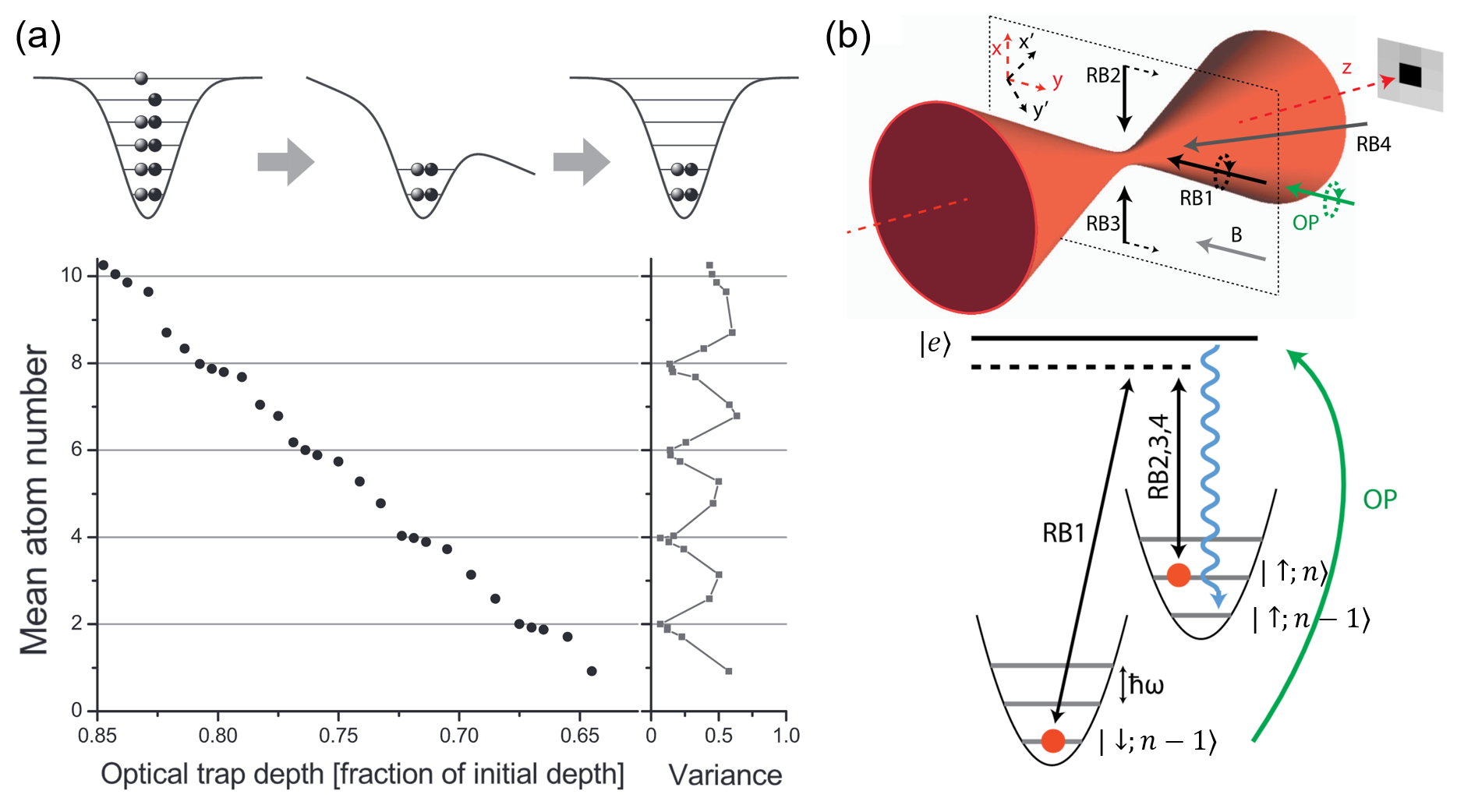}
	\caption{
		\textbf{(a)} The deterministic preparation of a few fermionic atoms in an optical tweezer. 
		The top figure shows the spill process. The excess atoms are spilled from the tweezer by adding a linear potential in the axial direction. By varying the trapping depth of the tweezer and the strength of the linear potential, the number of bound states in the trap can be controlled. After the atoms in the unbound states escape, the trap is restored, resulting in a few-particle quantum states with desired number of atoms. 
		The bottom figure shows the measured mean atom number (for about 190 measurements) as a function of the trapping depth, exhibiting a step-like dependence with plateaus for even atom numbers.  
		The variance figure shows that the number fluctuations are suppressed for even atom numbers.  
		\textbf{(b)} The experimental setup of a tweezer trap with sideband cooling and detection. 
		In the top figure, the atom is trapped at the waist of an optical tweezer created with a high-NA objective lens and is imaged along the $z$ axis with the same lens. The atom is cooled in three motional directions via sideband cooling with several Raman beams (RB). 
		The bottom figure shows the level diagram for the sideband cooling. A pair of Raman beams drive the coherent transition between $|\uparrow;n\rangle \to |\downarrow;n-1\rangle$. Then, the optical pumping (OP) recycles the atom back to the initial spin state via a spontaneous Raman process through an excited state ($|e\rangle$), where the spontaneously emitted photons carry away entropy.  After several repetitions of these steps, the atom is cooled to the motional ground state.  
		(a) Adapted with permission.\textsuperscript{\cite{Serwane2011}} Copyright 2011, American Association for the Advancement of Science.
		(b) Adapted with permission under the terms of the Creative Commons Attribution 3.0 License.\textsuperscript{\cite{Kaufman2012}} Copyright 2012, American Physical Society. 
	}
	\label{fig:Serwane2011Kaufman2012}
\end{figure*}

Later in 2011, Serwane et al.~\cite{Serwane2011} demonstrated a method of deterministic preparation of a few fermionic atoms into the ground state of an optical tweezer (see Figure~\ref{fig:Serwane2011Kaufman2012} (a)). 
The full control of the atom number and spin was achieved through a clever spilling technique. 
The technique of deterministic number preparation enabled significant experimental investigations, such as fermionization~\cite{Zurn2012}, the transition between few-body and many-body physics~\cite{Wenz2013}, quantum phase transitions~\cite{Bayha2020}, the formation of Pauli crystals~\cite{Holten2021}, and the study of few-body problems in an isolated environment~\cite{Reynolds2020, Weyland2021, Andersen2022}.

After loading the single atoms in the tweezers, the usual evaporative cooling method is not available as it depends on the collisions of many atoms. 
To further reduce the motional entropy in the tweezers, the atoms can be further cooled to their 3D motional ground state with the sideband cooling method~\cite{Kaufman2012, Thompson2013} (see Figure~\ref{fig:Serwane2011Kaufman2012} (b)).

The progress in these advanced techniques has led to the widespread applications of optical tweezers in various fields of quantum sciences and technologies. 
In the following, we will provide a concise overview of the development and diverse applications of tunnel-coupled optical tweezers.

\begin{figure*}[ht!]
	\centering
	\includegraphics[width=0.98\linewidth]{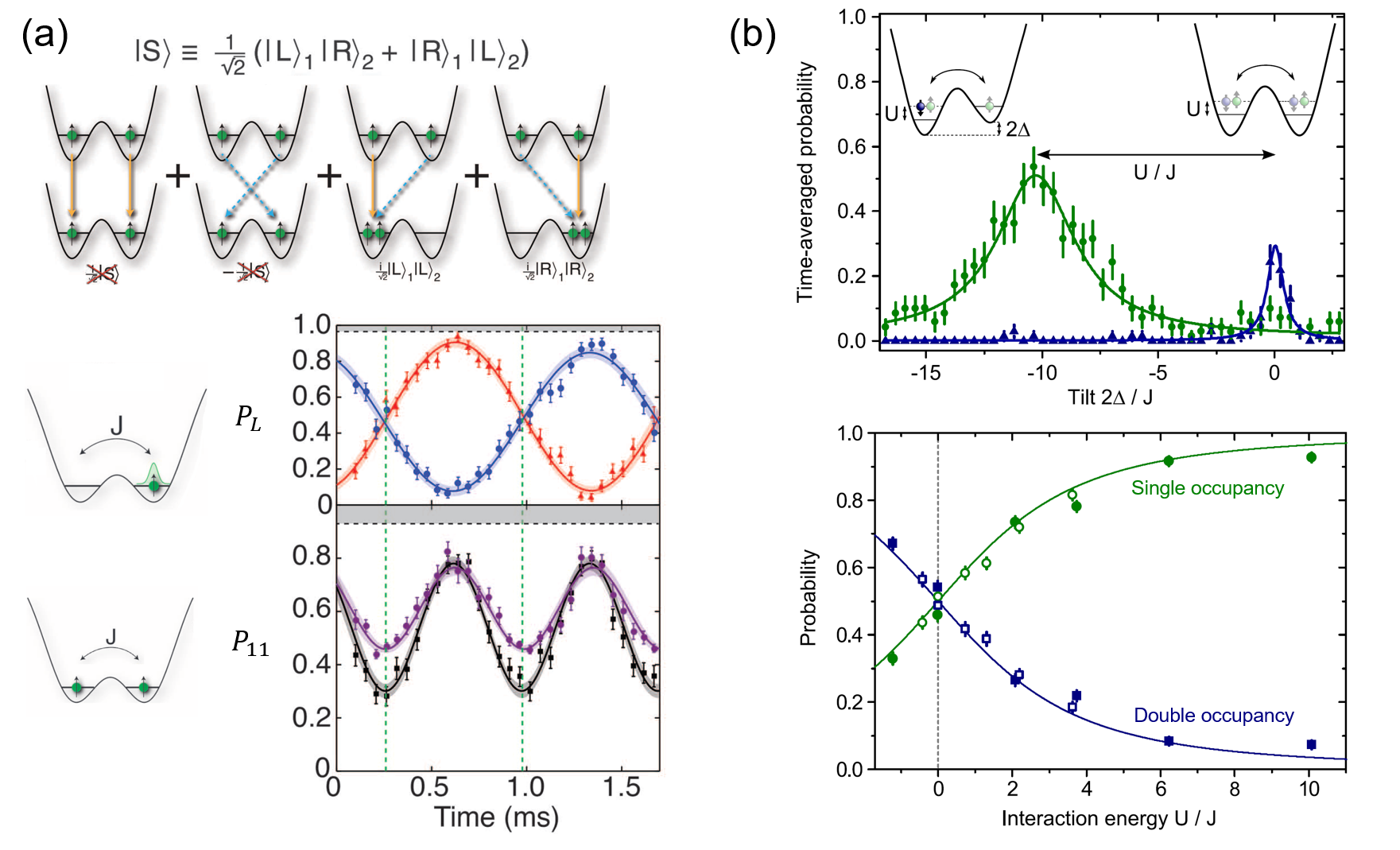}
	\caption{
		\textbf{(a)} The two-particle quantum interference and the observation of atomic Hong–Ou–Mandel effect. 
		The top figure shows the sketch of the tunneling dynamics of two bosonic $^{87}$Rb atoms in two tunnel-coupled optical tweezers. With perfect spin preparation and cooling to the ground state of each tweezer, the two atoms become indistinguishable. 
		During the tunneling of the two atoms, the two tweezers effectively act as an atomic beam splitter.  
		At the tunneling time when a balanced beam splitter is realized, the initial symmetric two-particle state $|S\rangle = (|L\rangle_1 |R\rangle_2 + |R\rangle_1 |L\rangle_2) / \sqrt{2}$ evolves into the doubly occupied state $\frac{\mi}{\sqrt{2}}(|L\rangle_1 |L\rangle_2 + |R\rangle_1 |R\rangle_2)$ due to the destructive interference. Here $|L\rangle$ ($|R\rangle$) denotes the lowest-energy single-particle state in the left (right) well, and the ket subscripts are the particle labels. 
		The bottom figures show the one-particle and two-particle tunneling dynamics of the atoms with tunneling energy $J$. 
		$P_L$ indicates the probability of finding one atom in the left well if one atom is initialized in the left (red) or right (blue) well. 
		$P_{11}$ indicates the probability of finding exactly one atom in each well if one atom is initialized in each well. 
		At the tunneling time when a balanced atomic beam splitter is realized (vertical green dashed lines), the measured results of $P_{11}$ (black) falls below the values for distinguishable particles (purple). 
		This indicates that the atoms are in a superposition of the doubly occupied state $\frac{\mi}{\sqrt{2}}(|L\rangle_1 |L\rangle_2 + |R\rangle_1 |R\rangle_2)$, a manifest of the Hong–Ou–Mandel effect. 
		\textbf{(b)} The two-site realization of Fermi-Hubbard model. The top figure shows the time-averaged probability of finding a single atom (green) or a pair of atoms (blue) in the right well as a function of the tilt $\Delta$ after two atoms are initialized in the left well. The interaction energy is about $U = 10.05\pm 0.19 J$ with $J$ the tunneling energy. The single-particle and pair tunneling are resonant (shown in the insets) at $\Delta =-U/2 $ and $\Delta =0$, respectively. 
		The bottom figure shows the relative probabilities  of finding two atoms in the same well (blue squares) or one atom in each well (green circles) for the ground state as a function of the interaction energy $U$. The measured results is in good agreement with the prediction of the Fermi-Hubbard model. For increasing repulsive interactions, the double occupancy is suppressed, indicating the crossover from a metallic to a Mott-insulating regime. For increasing attractive interactions, double occupancy is enhanced, indicating the onset of a charge-density-wave regime. 
		(a) Adapted with permission.\textsuperscript{~\cite{Kaufman2014}} Copyright 2014, American Association for the Advancement of Science. 
		(b) Adapted with permission.\textsuperscript{~\cite{Murmann2015}} Copyright 2015, American Physical Society.}
	\label{fig:Kaufman2014Murmann2015}
\end{figure*} 

\subsection{Individual atoms with indistinguishability }

\textbf{Hong-Ou-Mandel effect with atoms.} 
With the ability to cool single atoms to the 3D ground state of optical tweezers, the atoms can be made indistinguishable in a pair of tweezers. 
By experimentally studying the tunneling dynamics of two bosonic $^{87}$ Rb atoms cooled to the ground state of a tunnel-coupled pair of tweezers, Kaufman et al.~\cite{Kaufman2014} observed the Hong-Ou-Mandel effect of atoms (see Figure~\ref{fig:Kaufman2014Murmann2015} (a)). 
Initially, two atoms are prepared in the ground state of two tweezers, one per well. 
With perfect cooling and spin preparation, the two bosonic atoms are indistinguishable in the sense that all degrees of freedom other than their position (left or right well) are made the same. 
Then, the initial state can be denoted by $|S\rangle = \frac{1}{\sqrt{2}}(|L\rangle_1 |R\rangle_2 + |R\rangle_1 |L\rangle_2)$, where $|L\rangle$ ($|R\rangle$) is the lowest-energy localized state in the left (right) well and the ket subscripts are the particle labels. 
At a tunneling time $t=2\pi/8J$ with $J$ the tunnel coupling of a single atom, the pair of tweezers realizes a balanced atom beam splitter. 
Due to the destructive interference, the symmetric state evolves into the state $\frac{i}{\sqrt{2}}(|L\rangle_1 |L\rangle_2 + |R\rangle_1 |R\rangle_2)$, where two atoms occupy the same well at the same time. 
This is akin to the Hong-Ou-Mandel effect observed originally with photons~\cite{Hong1987}.

\textbf{Two-site Fermi-Hubbard model.}
With the ability of preparing few-particle systems in the ground state of optical tweezers, Murmann et al.~\cite{Murmann2015} prepared two fermionic $^6$Li atoms in two different hyperfine state in a tunnel-coupled two optical microtraps (see Figure~\ref{fig:Kaufman2014Murmann2015} (b)). 
With full control of the quantum system including the interparticle interaction strength, the tunneling rate, and the tilt of the potential, they realized the Mott-insulating and charge-density-wave states, and observed the influence of the second-order tunneling on the energy. 
With the realization of the building block of the Hubbard model, they demonstrated the feasibility of bottom-up approach of creating a quantum many-body system with single-site addressability.

\begin{figure*}[ht!]
	\centering
	\includegraphics[width=0.98\linewidth]{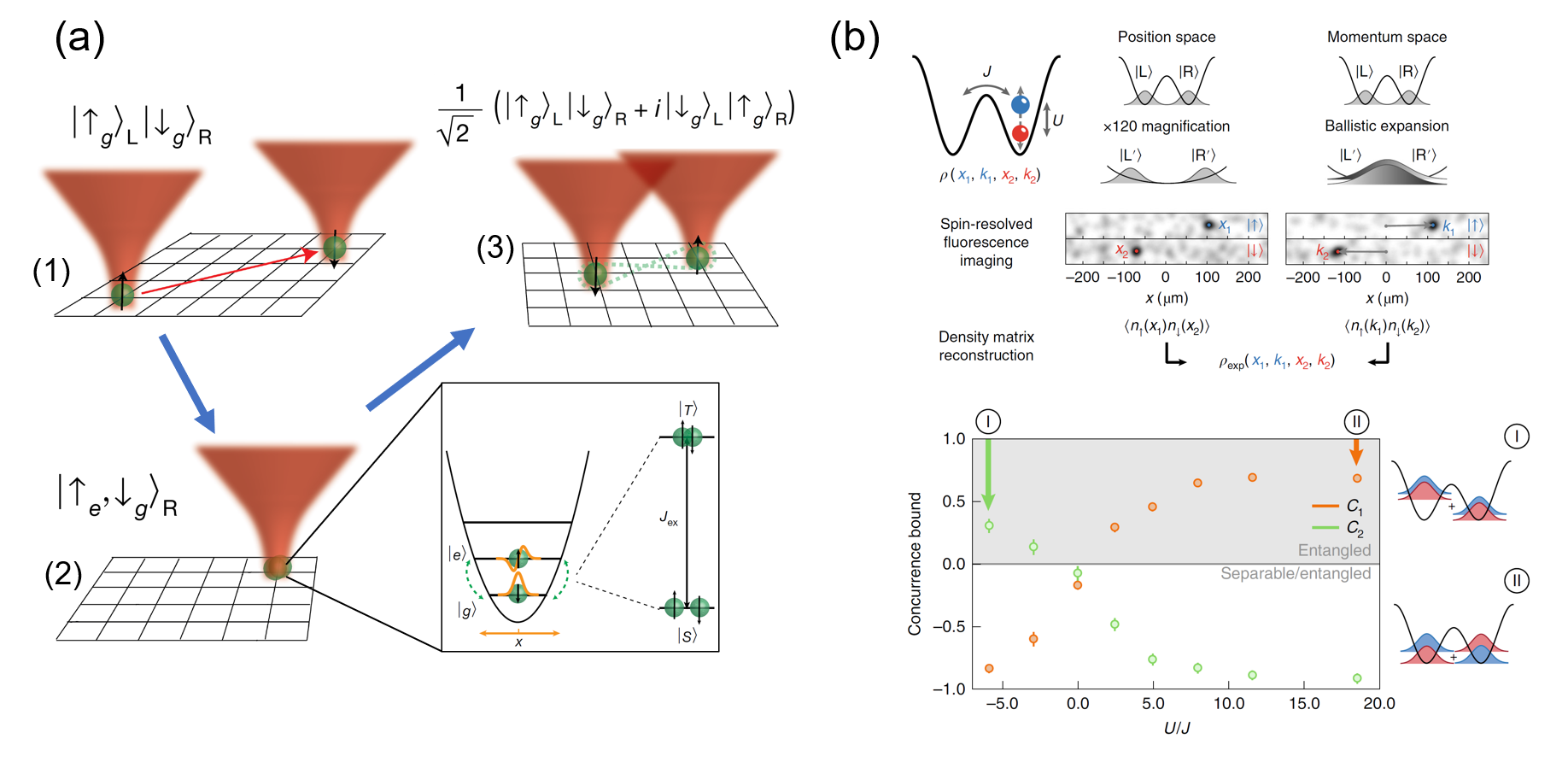}
	\caption{ 
		\textbf{(a)} The experimental sequence [(1) $\to$ (2) $\to$ (3)] for entangling two atoms via local spin-exchange interaction. Initially, in (1), two atoms (green) with opposite spins are prepared in the motional ground states of two separated optical tweezers (red). Here $g$ ($e$) labels the ground (excited) motional state, and $L$ ($R$) labels the left (right) tweezer. Then, in (2), the two tweezers merge and the two atoms are entangled via local spin-exchange interaction. The callout shows the sketch of the spin-exchange interaction with a rate $J_{ex}$. $|T\rangle$ and $|S\rangle$ denotes the spin-triplet and spin-singlet channels, respectively. Then, in (3), the two tweezers are separated and the preserved entanglement between the two atoms are detected. 
		\textbf{(b)} The top figures shows the sketch of experimental measurement of the single-particle spin-resolved correlations in position and momentum space for a two-site Fermi–Hubbard system. The correlation information can be used to reconstruct the density matrix of the initial state. The bottom figure shows that the result of the concurrence lower bounds $C_1$ and $C_2$~\cite{Bergschneider2019}, extracted from the measured correlations, certify entanglement in gray region. The spin up and down components are sketched in blue and red in the side panels. 
		(a) Adapted with permission.\textsuperscript{~\cite{Kaufman2015}} Copyright 2015, Springer Nature. 
		(b) Adapted with permission.\textsuperscript{~\cite{Bergschneider2019}} Copyright 2019, Springer Nature.
	}
	\label{fig:Kaufman2015Bergschneider2019}
\end{figure*}

\textbf{Quantum entanglement and correlation.}
Later, using spin-exchange interactions and tunnel coupling in two mobile optical tweezers, people have demonstrated locally entangling two ultracold neutral atoms and separating them without destroying the entanglement~\cite{Kaufman2015}, and this technique provided a framework for dynamically entangling remote qubits via local operations in quantum computations (see Figure~\ref{fig:Kaufman2015Bergschneider2019} (a)). 
Furthermore, Lester et al.~\cite{Lester2018} used atomic Hong-Ou-Mandel interference to postselect a spin-singlet state after overlapping two atoms in distinct spin states on an effective beam splitter in double wells formed by tunnel-coupled optical tweezers.

Bergschneider et al.~\cite{Bergschneider2019} measured the two-body correlations in position and momentum space of two $^6$Li atoms with different spins in a two-site Fermi-hubbard optical tweezers (see Figure~\ref{fig:Kaufman2015Bergschneider2019} (b)). 
Via the single-particle resolved measurements of atomic momenta, 
they obtained the coherence properties and quantified different types of entanglement, such as the entanglement between the spatial modes or the spin modes. 
From the correlation information, they also showed the reconstruction of the density matrices and the R\`enyi entanglement entropies for different interaction strengths (attractive or repulsive), which gives a comprehensive characterization of the quantum states.

Later, through measurements on a few-fermion system in a multiwell potential formed by a few optical tweezers, Becher et al.~\cite{Becher2020} studied the influence of fermionic exchange antisymmetry on the indistinguishable particle entanglement.  
Preiss et al.~\cite{Preiss2019} studied the two- and three-body correlations of a few $^6$Li atoms in optical tweezers. 
The fermionic atoms become highly indistinguishable when cooled into the ground states of and the interference of pairs of atoms with high contrast (close to $80\%$) was found~\cite{Preiss2019}.

\begin{figure*}[ht!]
	\includegraphics[width=0.95\linewidth]{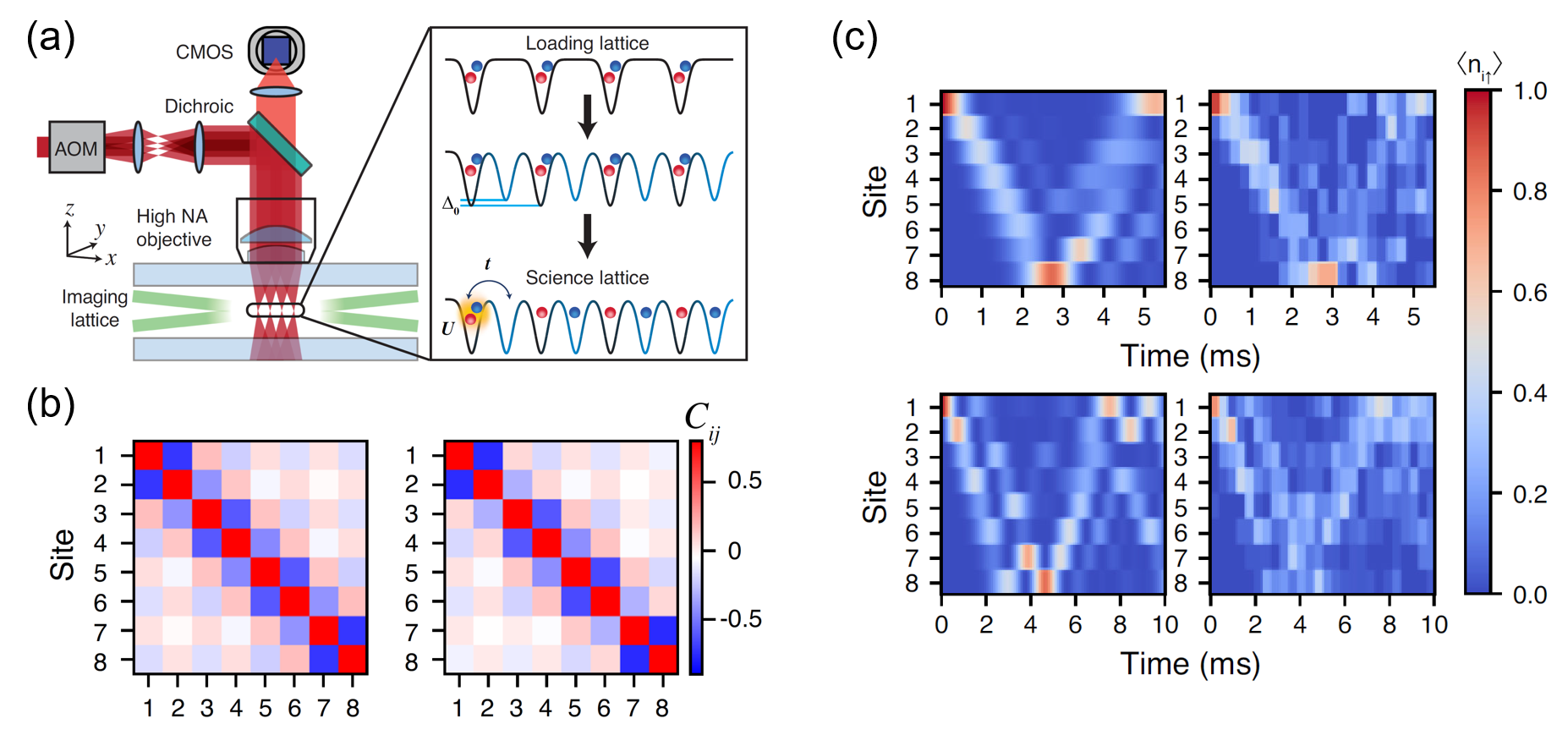}
	\caption{ 
		\textbf{(a)} Experimental setup for the generation of a Fermi-Hubbard chain in a 1D tunnel-coupled optical tweezer array. Initially, four atoms per spin state are loaded in the loading lattice. Then four additional tweezers with an energy offset $\Delta_0$ are ramped on, changing the filling of the system to $1/2$. The Fermi-Hubbard array is formed by reducing $\Delta_0$ to zero. 
		\textbf{(b)} The up-up antiferromagnetic correlations between all lattice sites. The left and right figures show the theoretical calculation from exact diagonalization at temperature $T = 0.21(3) t / k_B$ and the experimentally measured result for $U/t = 6.7(3)$, respectively. 
		\textbf{(c)} The tunneling dynamics of a pair of noninteracting atoms initialized on the edge site. Figures in the top and bottom rows shows the tunneling dynamics in a configuration with equal tunneling rates ($t = 296(3){\rm Hz}$) and staggered tunneling rates ($t_{i,i+1} = 289(3){\rm Hz} $ for odd $i$, or $t_{i,i+1} = 213(2){\rm Hz}  $ for odd $i$), respectively. The left and right columns show the exact diagonalization and experimental results, respectively. 
		Adapted with permission.\textsuperscript{\cite{Spar2022}} Copyright 2022, American Physical Society. }
	\label{fig:Spar2022}
\end{figure*} 

\subsection{Tunnel-coupled optical tweezer array}

\textbf{1D Fermi-Hubbard optical tweezer array.}
Optical lattices are well-established efficient way of creating periodic trapping potential with up to more than thousands of lattice sites. 
This works as a top-down approach, in which quantum states could be prepared by directly loading the ultracold quantum gases into the optical lattices. 
Combined with quantum gas microscopes~\cite{Bakr2009, Sherson2010, Cheuk2015, Parsons2015, Haller2015, Edge2015, Gross2021} which provide atom imaging with single-site resolution, the phase diagram of Fermi-Hubbard model could be probed on the optical lattices. 
However, the approach of studying Fermi-Hubbard model using optical lattices encounters two major challenges. 
First, the correlated states in optical lattices have relatively large entropy per particle. 
The lowest entropies achieved is about $0.3 - 0.5 k_B$ per particle with $k_B$ the Boltzmann constant~\cite{Mazurenko2017, Brown2019, Sompet2022}. 
The relatively large entropies make it difficult to access the interesting regime of the phase diagram of the square Hubbard model like the pseudogap or the $d$-wave superconductor. 
Second, it is difficult to reconfigure the apparatus to study different lattice geometries. 
These challenges could be potentially well addressed by the Fermi-Hubbard tweezer arrays. 
The tweezer arrays allow precise dynamical control of model parameters such as the lattice geometries, energy offsets, the tunneling matrix element at the single site level.

Recently, Spar et al.~\cite{Spar2022} demonstrated the preparation of an eight-site 1D Fermi-Hubbard chain with $^6$Li atoms on a tunnel-coupled optical tweezer array (see Figure~\ref{fig:Spar2022}). 
The experimental setup is shown in Figure~\ref{fig:Spar2022} (a).
Compared to other alkali atoms, the light-mass $^6$Li atoms could provide relatively large tunnel coupling between the adjacent sites.
Initially, the $^6$Li atoms on the two hyperfine spin states are prepared in the ground state of four independent tweezers. 
By ramping on four additional tunnel-coupled tweezers, a correlated state with nearly half filling is adiabatically obtained (See the experimental procedure in Figure~\ref{fig:Spar2022}). 
The typical tunneling energies and tweezer depths are about $h \times 200 {\rm Hz}$ and about $h \times 50 {\rm kHz}$, respectively. 
To obtain a correlated state, the disorder of tweezer depths need to be smaller than $0.5\%$.  
This condition is more stringent that the tweezer arrays with Rydberg atoms, which only requires a few percent~\cite{Young2020}.

With a quantum gas microscope, single-site detection is achieved and Mott insulators with strong antiferromagnetic correlations are observed. 
Through the spin correlations, the upper bound of the entropy per atom is found to be about $0.26 k_B$, which is comparable to the lowest entropies achieved with optical lattices. 
With full postselection on the atom number and spin, the entropy from the initial state can be further eliminated and the total entropy of the states are limited only by adiabaticity of the preparation. 
Moreover, with the precise control up to the single site level on the tweezer platform, after initializing atoms on one tweezer, the tunneling dynamics  across the array for uniform and staggered 1D geometries are observed. 
This platform is suitable for studying ground states of many-body systems at high $U/t$ and at half-filling, where residual disorder is relatively small. 
The system can be further extended to arrays of up to a hundred tweezers, as demonstrated by the Rydberg atom arrays, and could also be scaled up to 2D tweezer arrays.

\begin{figure*}[ht!]
	\includegraphics[width=0.95\linewidth]{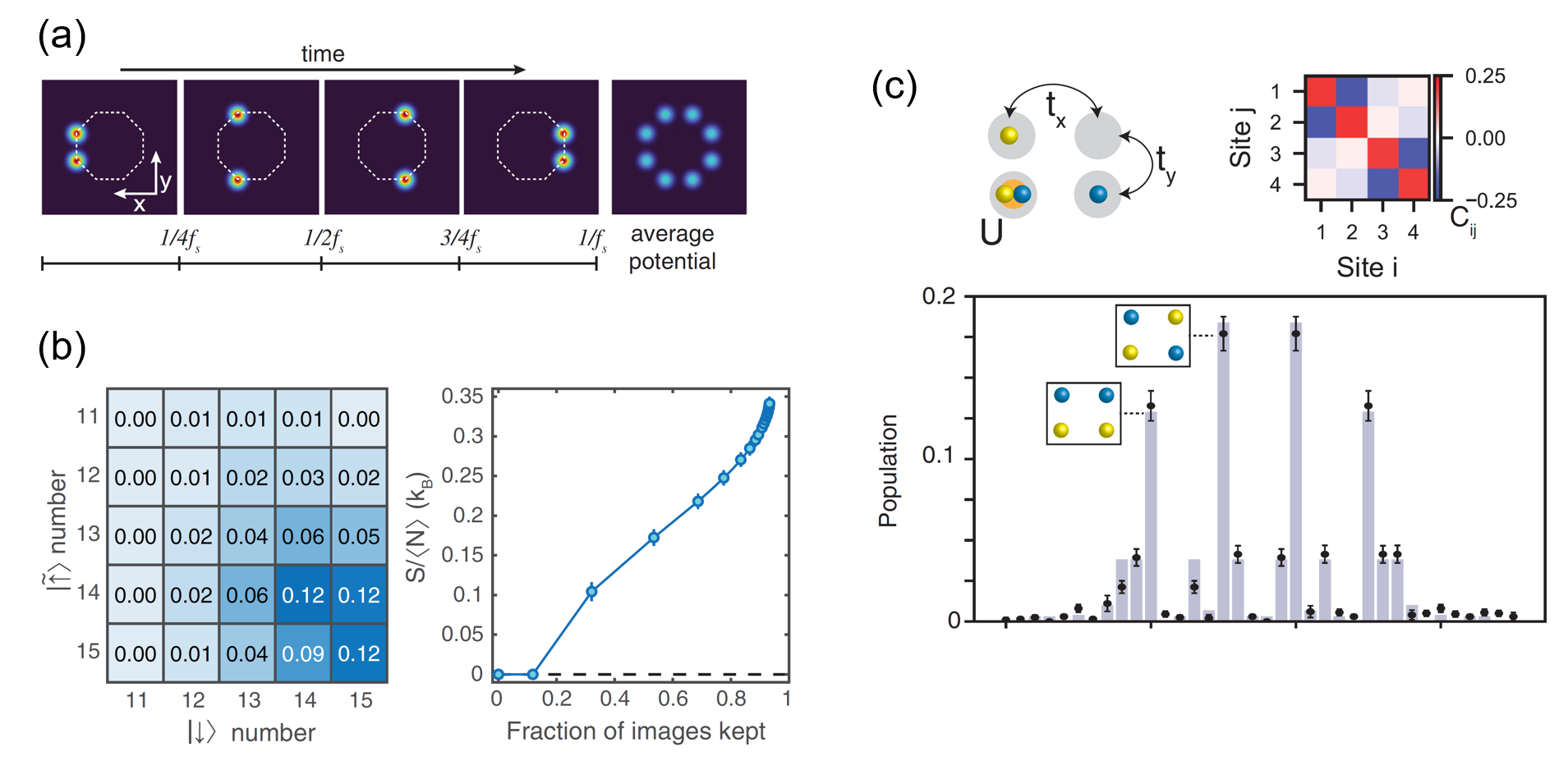}
	\caption{
		\textbf{(a)} 2D stroboscopic tweezer array. The four figures on the left show the time sequence of stroboscopic array generation of an eight-site ring. Each column of the array is switched on for a quarter of the strobe period $1/f_s$ with $f_s$ the strobe frequency. The tweezer array is given by the time-averaged potential (the fifth figure).  
		\textbf{(b)} The left figure shows the probability distribution (over 972 shots) as a function of the number of atoms in each spin state for a $3\times5$ rectangular array. The right figure displays the effective entropy as a function of the fraction of images kept after postselecting. This shows that the the entropy can be reduced to zero in principle through postselection. About $12\%$ of the images had perfect filling, giving a zero entropy. Keeping images with up to two holes (over $50\%$ of the shots) gives a low entropy of $\approx 0.17 k_B $ per particle. 
		\textbf{(c)} The low-entropy preparation of a 2D Fermi-Hubbard
		model in a tunnel-coupled $2\times 2$ tweezer array. 
		The top left figure shows the sketch of the $2\times 2$ array with the Fermi-Hubbard model parameters, the tunneling matrix elements $t_x/h = 140(5){\rm Hz}$, $t_y/h = 220(5){\rm Hz}$ and the on-site interaction $U/\bar{t} = 3.4(2)$ with $\bar{t} = (t_x+t_y)/2$.
		The top right figure shows the spin-spin correlations $C_{ij} = \langle S_{z,i} S_{z,j} \rangle - \langle S_{z,i} \rangle \langle S_{z,j} \rangle$ where $S_{z,i} = (n_{\uparrow,i} - n_{\downarrow,i})/2$ with $n_{\uparrow,i}$ ($n_{\downarrow,i}$) the atom number of spin up (down) at site $i$. 
		The bottom figure shows the measured population in each microstate (black dots). The best fit (purple bars) gives an entropy $\le 0.09 k_B$ per particle. The two insets show the two most common microstates.  
		Adapted with permission.\textsuperscript{\cite{Yan2022}} Copyright 2022, American Physical Society. } 
	\label{fig:Yan2022}
\end{figure*}

\textbf{2D Fermi-Hubbard optical tweezer array.}
Yan et al.~\cite{Yan2022} realized a 2D array of $^6$Li atoms on tunnel-coupled optical tweezer arrays (see Figure~\ref{fig:Yan2022}). 
They demonstrated the ability of forming 2D arrays with various geometries. 
For tunnel-coupled arrays, the inhomogeneity of tweezer depths must be controlled to be at least less than the tunneling energies, or typically better than $0.5\%$ of tweezer depths. 
To achieve the homogeneity requirement, the authors used a stroboscopic tweezer technique. 
The array is generated one column at a time. 
When the strobe rate is much larger than the tweezers' harmonic trapping frequencies, the atoms experience a time-averaged potential of the 2D array. 
Then, highly homogeneous 2D arrays with nearly arbitrary geometry could be generated using this stroboscopic technique. 
In their article, different geometries such as $5\times5$ rectangular array,  21-site Lieb plaquette, $4\times5$ triangular array, and octagonal ring arrays were demonstrated. 
Note that recently the calculation methods for the Hubbard parameters such as the tunneling energy, the on-site potential, and the on-site interaction for tunnel-coupled tweezer arrays with arbitrary geometries have been developed~\cite{Wei2023}.

In the experiment, the prepared initial state have low entropy about $0.34k_B$ per particle, comparable to the lowest achieved entropies (about $0.25 - 0.5 k_B$ per particle) in the state-of-the-art optical lattice experiments~\cite{Brown2019, Chiu2018, Mazurenko2017, Sompet2022}. 
Moreover, by post-selecting the images with high filling, the initial state entropy can be further reduced and could be reduced to zero if only selecting the images with the population per spin equal to the number of tweezers. 
Then, the total entropy would be mainly limited by the adiabatic ramping from the initial states to the correlated states.

Then, the simplest 2D Fermi-Hubbard model was implemented in a tunnel-coupled $2\times 2$ optical tweezer array~\cite{Yan2022}. 
The spin-spin correlations were measured by a bilayer imaging scheme. 
For postselected experimental shots, the total entropy was found to be less than $0.09k_B$ per particle. 
Through the preparation of a correlated state in the two-by-two tunnel-coupled Hubbard plaquette, this work demonstrated the building blocks for realizing a programmable fermionic quantum simulator. 
It also showed the advantages of the tunnel-coupled optical tweezer array platform: the flexibility in forming various geometries and the low entropy of preparation.

\begin{figure*}[ht!]
	\includegraphics[width=0.95\linewidth]{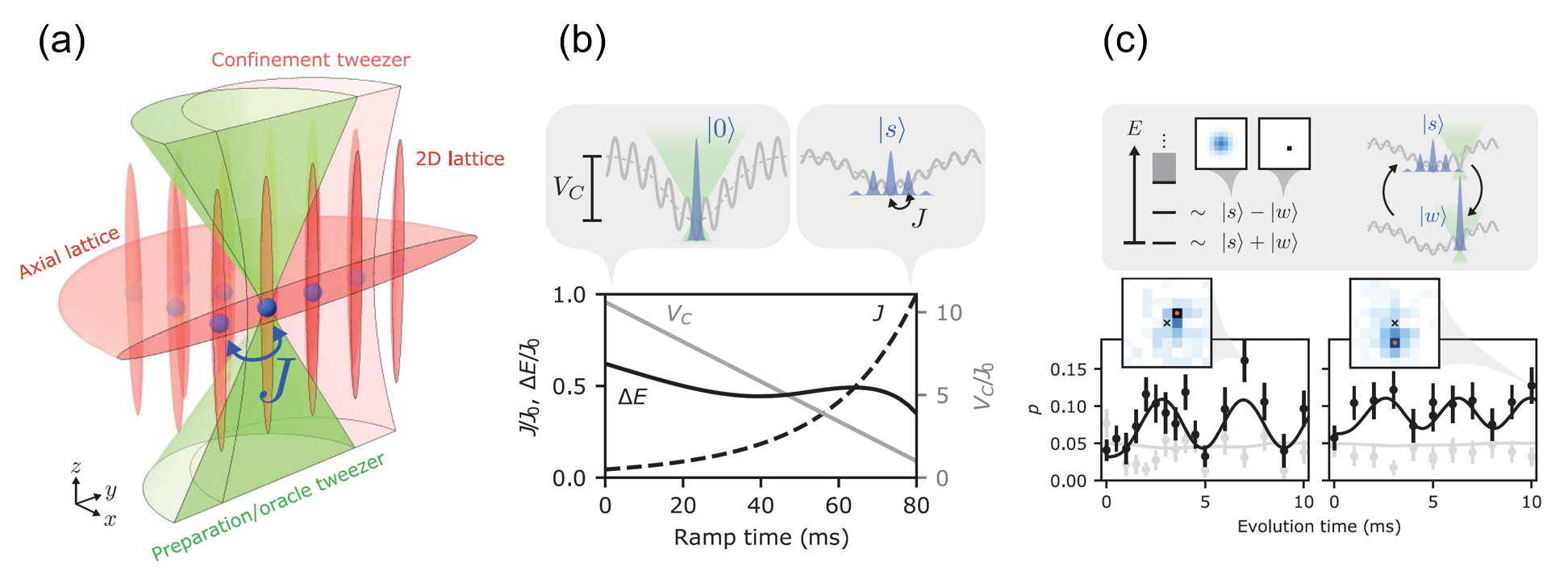}
	\caption{
		\textbf{(a)} Experimental setup for 2D quantum walks with the integration of optical tweezers and optical lattices. The preparation tweezer (green) is used to implant individual $^{88}$Sr atoms (solid blue sphere) into the 2D optical lattice (red tubes). The atoms can move through the tunnel-coupled lattice with a tunneling energy $J$. The preparation tweezers can further be used to modify the depth of individual sites. A confinement tweezer with large waist (pink) can be used to add a tunable harmonic potential spanning over many sites. 
		\textbf{(b)} The adiabatic prepration of the resource state $|s\rangle$. Initially, through a confinement tweezer (green), an atom is implanted in the lowest energy state $|0\rangle$ in a deep lattice with no tunneling. Then, $|0\rangle$ is adiabatically connected to $|s\rangle$ through the ramping of the tunneling energy $J$ and the depth $V_C$ of the confinement tweezer, while keeping the energy gap $\Delta E$ between the ground state and the first excited state roughly fixed. $J$, $V_C$, and $\Delta E$ are shown in the bottom figure.
		\textbf{(c)} Demonstrations of spatial search algorithm. An oracle Hamiltonian $H_w = -V_w |w\rangle \langle w|$ is applied through a tweezer (green). The ground and first excited states become approximately the even and odd superposition of the arbitrary marked site $|w\rangle$ and the the resource state $|s\rangle$, respectively. The evolution of the initial state $|s\rangle$ under $H_w$ leads to the coherent oscillations between $|s\rangle$ and $|w\rangle$. In the bottom figures, for different positions of oracles (red points in the insets, with the lattice center marked by a cross), the measurements (black points) of the oscillations in the population on the marked site $|w\rangle$ is in good agreement with theory (black curves). For comparison, the populations in $|0\rangle$ at the lattice center is also shown (gray points and theory curves).
		Adapted with permission.\textsuperscript{\cite{Young2022}} Copyright 2022, American Association for the Advancement of Science. } 
	\label{fig:Young2022}
\end{figure*} 

\subsection{Novel applications and advances}

\textbf{Quantum walks with the integration of optical tweezers and optical lattices.} 
Optical tweezers and optical lattices each possess distinct advantageous characteristics. 
Optical tweezers excel in their ability of programmable control of atoms at the single-site level. 
On the other hand, optical lattices offer a more scalable and uniform periodic trapping potential with stronger tunnel coupling. 
Recognizing the benefits of both techniques, Young et al.~\cite{Young2022} have successfully combined these tools, harnessing their favorable properties to achieve programmable 2D quantum walks(see Figure~\ref{fig:Young2022}).

They first trapped and cooled individual $^{88}$Sr atoms in optical tweezers. 
Then the tweezers can be used to implant the atoms into one 2D layer of a 3D optical lattices. 
Due to the tunnel-coupling of neighboring sites, the implanted atoms undergo continuous-time quantum walks in 2D. 
Then, the spatial search quantum algorithm was demonstrated. 
In the presence of a confinement tweezer and through the implantation tweezer, an atom can initially be placed in the site with the lowest energy in a deep lattice. 
Then, by adiabatic ramping up the tunneling energy and down the confinement tweezer, the initial state is connected to the desired resource state, which is the ground state $|s\rangle$ of the lattice Hamiltonion
\begin{equation}
	H_{\rm lat} = - J \sum_{ij}A_{ij}|i\rangle \langle j| - \sum_i  V_i|i\rangle \langle i|,
\end{equation}
where $J$ is the tunneling energy and $V_i$ is the local energy shift. 
$A_{ij}=1$ if the site $i$ and $j$ are connected, or $A_{ij}=0$ if not. 
The resource state $|s\rangle$ is reported be prepared with a fidelity about $76\%$. 

The quantum-walk-based search algorithms can be realized by an extra oracle Hamiltonion $H_w = -V_w |w\rangle \langle w|$, which can be realized through a tweezer with appropriate strength.  
The quench to the search Hamiltonion $H=H_{\rm lat} + H_w$ leads to coherent oscillation between the resource state $|s\rangle$ and $|w\rangle$. 
After a half period of this oscillation, measuring the position of the walker will identify the marked site $|w\rangle$. 
The authors performed a proof-of-principle demonstration of spatial search on a novel platform, which combines the single-site programmability of optical tweezer array and clean environment for tunneling and scalability of optical lattice.

Recently, Young et al.~{\cite{Young2023}} showcased a groundbreaking atomic boson sampler involving as many as 180 atoms, achieved by combining optical tweezers and tunnel-coupled optical lattices. Optical tweezers facilitate rapid and programmable assembly of atom ensembles through rearrangement, while tunnel-coupled optical lattices enable flexible temporal evolution of quantum many-body states and high-fidelity atom detection. This work introduces a fresh approach to leveraging the tweezer-lattice synergy for quantum simulation and computation.

\textbf{Other recent advances.}
The tunnel-coupled optical tweezer platform is actively being explored for innovative applications.
Florshaim et al.~\cite{Florshaim2023} successfully demonstrated the implementation of spatial adiabatic passage for ultracold atoms using tunnel-coupled optical tweezers, showcasing the advanced level of control and manipulation achievable in the optical tweezer array platform. 
Moreover, Gonz\'alez-Cuadra et al.~\cite{Gonzalez-Cuadra2023} proposed to build fermionic quantum processors based on the non-local tunneling gates to perform quantum phase estimation and simulate lattice gauge theory dynamics. 
The proposed fermionic quantum processors could be implemented through the tunnel-coupled pairs of tweezers worked as fermionic beam splitters for the fermionic atoms cooled to the ground state of the optical tweezers.

\textbf{Emerging optical field manipulation techniques for ultracold atoms.} 
The emergence of optical field manipulation techniques, such as optical superoscillation and metasurfaces, holds great promise for advancing the generation of optical tweezers with smaller trapping sizes, larger tunnel coupling strengths, and more compact experimental apparatus.
Recently, Rivy et al.~\cite{Rivy2022} demonstrated the trapping of single atoms in a superoscillatory optical trap, which features a much smaller trap size, reaching a subwavelength radius, potentially allowing for the engineering of larger tweezer arrays and stronger tunnel coupling due to the reduced spacing.

Moreover, optical metasurfaces, composed of subwavelength dielectric or metallic nanostructures, provide the comprehensive manipulation of the amplitude, phase, and polarization of light, by designing the nanostructure shape, size, and material composition~\cite{Kamali2018, Chen2020, Liu2021, Solntsev2021, Luo2023}. 
In a recent study by Hsu et al.~\cite{Hsu2022}, they successfully demonstrated the trapping and imaging single $^{87}$Rb atoms in optical tweezers generated by a highly efficient dielectric metasurface lens, expanding the use of metasurfaces in optical dipole traps and further complex quantum information experiments with neutral atoms. 
Also, dielectric metasurfaces have been demonstrated to generate a single-beam magneto-optical trap for the trapping $\approx 10^8$ cold $^{87}$Rb atoms by splitting a single incident beam into five separate beams with defined polarization and uniform energy distributions~\cite{Zhu2020, Jin2023}, enabling the creation of a compact cold atom source apparatus.
Recently, a variety of optical trap arrays have been demonstrated~\cite{Huang2023} via metasurface holograms, featuring high positioning accuracy, size uniformity, optical efficiency, and thermal stability and allowing the generation of up to thousands of trap sites with dense spacing in a compact system.
Ongoing research is actively exploring novel applications of metasurfaces in quantum sciences and technologies, and aiming to unlock new frontiers and possibilities in the fields~\cite{Liu2021, Solntsev2021, Luo2023}.

Furthermore, the emerging techniques of vector optical field manipulation provide additional knobs for controlling over the polarization states of light in both spatial and temporal domains. 
Vector optical fields exhibit a nonuniform polarization distribution on the wavefront, or in a more general sense, nonuniform distribution of electric and magnetic field vectors on a surface in 4D spacetime. 
Notably, polarization gradients, a type of vector optical field, are widely employed for sub-Doppler cooling of ultracold atoms. 
Recent advancements in the manipulation of vector optical fields~\cite{Luo2022a, Zhang2022, Zhao2023} are offering new possibilities for enhanced control of ultracold atoms in 4D spacetime.

\section{Summary and outlook}
\label{sect:summary}

In summary, we have provided an overview of the development and recent progress in tunnel-coupled optical microtraps, ranging from early studies on superfluids in double-well systems to the more recent trapping of individual atoms in optical tweezer arrays. 
We have discussed the tunneling dynamics and atomic Josephson effect of superfluids in the double-well systems. 
Subsequently, we have explored the development of optical tweezers, which are micrometer-scale traps created using tightly focused beams of light generated with a high-NA objective lens, for the trapping of individual atoms. 
The cooling of individual atoms to the motional ground state allows for the observation of phenomena such as the atomic Hong-Ou-Mandel effect arising from the indistinguishability of bosonic atoms, as well as the realization of miniature two-site Fermi-Hubbard models, and also facilitates the study of quantum entanglement and correlations. 
Furthermore, we have reviewed recent advances in the realization of 1D Fermi-Hubbard chains and 2D Fermi-Hubbard arrays with stroboscopic technique. 
Lastly, we have introduced the novel application of optical tweezers in quantum walks, achieved through their integration with optical lattices.

One important future direction involves increasing the array size of 2D tunnel-coupled tweezers arrays for itinerant atoms, allowing quantum simulation of actual many-body systems such as multileg triangular ladder systems, quantum spin liquids, and systems with flat bands and Dirac points. 
The challenge along this direction is achieving low disorder of tweezer depths, dense spacing, and therefore sufficiently strong tunnel coupling in large arrays. 
This might be accomplished by the utilization of novel techniques such as the stroboscopic tweezer arrays~\cite{Yan2022}, the optical superoscillation~\cite{Rivy2022}, or more homogeneous arrays produced by metasurfaces~\cite{Hsu2022, Huang2023}. 
On the other hand, tunnel-coupled optical tweezers offer a favorable approach to studying many-body ground states with a large interaction-tunneling ratio, where the disorder of tweezer depths becomes negligible, opening up exciting possibilities for exploring phenomena such as Nagaoka ferromagnetism~\cite{Nagaoka1966}.

Another promising avenue for future research is the integration of optical tweezers and optical lattices, following the philosophy outlined in Ref.~\cite{Young2022}. By harnessing the precise programmable control of optical tweezers at the single-site level, it becomes feasible to precisely engineer the quantum many-body states in large-size optical lattices and subsequently study the quantum many-body dynamics with strong tunnel coupling in the lattices. 
This advancement would enhance the investigation of quantum many-body dynamics in optical lattice systems, offering new insights and opportunities for future experimental exploration.

\begin{acknowledgments}
	S.Z. gratefully acknowledged support from the Youth Innovation Promotion Association, Chinese Academy of Sciences (2023399).  
\end{acknowledgments}


\end{document}